\documentclass[final,3p,times,twocolumn]{elsarticle}

\usepackage{graphicx}
\usepackage{amssymb,amsmath}

\newcommand{\dcrit}{3\sqrt{3}M}
\newcommand{\chiin}{\chi^{\rm in}}
\newcommand{\chiout}{\chi^{\rm out}}
\newcommand{\rin}{r^{\rm in}}
\newcommand{\rout}{r^{\rm out}}
\journal{Physics Letters B}

\begin{document}

\begin{frontmatter}

\title{Apparent motion of a spherical shell collapsing onto a black hole} 
\author{Robert F. Penna}
\ead{rpenna@cfa.harvard.edu}

\address{Harvard-Smithsonian Center for Astrophysics,
60 Garden Street, Cambridge, MA 02138, USA }

\begin{abstract}

We model the collapse of a spherically symmetric, constant density,
pressureless shell onto a preexisting black hole.  Contrary to a
recent claim of Liu and Zhang (2009), we show that an observer at
infinity never sees any part of the shell cross the event horizon.
The entire shell appears to ``freeze'' outside the black hole.  We
find that during intermediate stages of the collapse, the change in
apparent size of the inner surface is non-monotonic.  However, at late
times, the apparent sizes of the inner and outer surfaces of
the shell both approach $3\sqrt{3}GM/c^2$ asymptotically, where $M$ is
the combined mass of the initial black hole and the shell.

\end{abstract}


\end{frontmatter}

\section{Introduction}
\label{sec:intro}

The collapse of a spherically symmetric, constant density,
pressureless ``star'' to a black hole was solved by Oppenheimer and
Snyder \cite{OS39} (hereafter OS39).  They showed that an observer on
the star hits a singularity in finite proper time, while an observer
at infinity sees the star slow down and ``freeze'' as it approaches
the event horizon.  The apparently contradictory experiences of the
two observers was not reconciled until Finkelstein found a
horizon-penetrating coordinate system accommodating both viewpoints
\cite{fink58}.

The OS39 model has since been generalized in many directions.  Misner and
Sharp \cite{misner64} discussed adiabatic collapse of an ideal
fluid.  Vaidya added radiation and gave non-adiabatic solutions
\cite{vaidya66}.  Planar and cylindrically-symmetric collapse
scenarios were considered by Liang \cite{liang74}.  The collapse of a
slowly rotating cloud to a Kerr black hole was modeled by Kegeles
\cite{kegeles78}.  Magnetized collapse was investigated
numerically by Baumgarte and Shapiro \cite{baum03}.

None of these generalizations challenged the original conclusion that
the observer at infinity sees the collapsing material freeze as it
approaches the horizon.  However, recently Liu and Zhang \cite{LZ09}
(hereafter LZ09) claimed that when a finite thickness shell collapses
onto a preexisting black hole, all but an infinitesimal surface layer
of the shell disappears across the horizon in finite time.  Their
argument rests on the construction of a time coordinate in the shell
that they identify with the observer at infinity's proper time.  We
disagree with this interpretation.  If one computes the light travel
time across the shell in their coordinates, one finds it diverges as
the inner surface of the shell approaches the horizon.  So no part of
the shell is observed entering the hole and it freezes outside.

In addition to potential time delay as a result of gravitational
redshift, there is also a geometrical time delay.  Light rays in the
Schwarzschild metric with impact parameters near $D=\dcrit$ (where we
set $G=c=1$) follow tightly wound spirals.  As $D\rightarrow \dcrit$,
the path length increases and the time delay diverges.  So the outer
edge of a collapsing star's disk appears to freeze away from the
horizon, at $3\sqrt{3}M$ \cite{ames68}.

In this Letter, we give a solution for the collapse of a shell onto a
preexisting black hole that is simpler than the construction of LZ09.
We describe how the horizon expands with time and show that the light
crossing time across the shell diverges as the material approaches the
horizon.  We solve for the apparent sizes of the inner and outer
surfaces of the shell as a function of time.  At intermediate stages
of the collapse, the apparent size of the inner surface 
varies non-monotonically.  At late times, both surfaces freeze as
they approach size $3\sqrt{3}M$, where $M$ is the combined mass of the
initial black hole and the shell.

\section{Collapse solution}

\label{sec:collapsesoln}

\subsection{Metric in the exterior of the shell}

Consider a spherical shell with negligible internal pressure which
collapses onto a preexisting black hole.  The geometry of the
spacetime external to the shell will be the Schwarzschild geometry
throughout the collapse, by Birkhoff's theorem.  This is true of any
spherical configuration which does not radiate an appreciable fraction
of its mass.  General relativity prohibits monopole gravitational
waves, so there is no possible way for any gravitational influence of
the radial collapse to propagate outward \cite{thorne67}.  By the same
argument, we also have the Schwarzschild geometry in the cavity
between the shell and black hole.  The coordinates constructed by LZ09
are not Schwarzschild, but this is an artifact of their construction
and not, as they claim, a failure of Birkhoff's theorem.

\subsection{Metric in the interior of the shell}

First consider the interior of a collapsing, constant density star.  The metric 
is a patch of the Friedmann-Robertson-Walker (FRW) metric :
\begin{align}\label{eq:frw}
ds^2 &= a^2(\eta) \left[-d\eta^2+d\chi^2+\sin^2\chi d\tilde{\Omega}\right],\\
d\tilde{\Omega} &= d\tilde{\theta}^2+\sin^2\tilde{\theta} d\tilde{\phi}^2.
\end{align}
This is expected because the spacetime in the star is homogeneous and
isotropic.  It can also be derived from Einstein's equations
\cite{weinberg}.  The solution used in cosmology to describe a
spatially flat universe has curvature $K=0$, but the interior of the
star has the spatial geometry of a 3-sphere, so $K=1$.  The Einstein
equations demand for the scale factor
\begin{equation}
a\left(\eta\right)=\left(a_i/2\right)\left(1+\cos\eta\right).
\end{equation}

We let the time coordinate $\eta$ run from $0$ to $\pi$, over which
period the FRW universe contracts from an initial size $a=a_i$ to a
singularity.  We limit the radial coordinate to the range $0<\chi
\leq \chiout$, and match onto the external Schwarzschild geometry at
$\chiout\equiv\sin^{-1}\left(\rout_i/a_i\right)$.  The angular
coordinates $\left(\tilde{\theta},\tilde{\phi}\right)$ have their usual ranges and can
be matched with the angular coordinates of the Schwarzschild
metric at the boundary of the star.  The resulting star has mass $M=\left(a_i/2\right) \sin^3
\chiout$, initial radius $\rout_i=a_i \sin\chiout$, and
initial density
\begin{equation}\label{eq:rhoi}
\rho_i=\frac{3}{8\pi a_i^2}=\frac{M}{4/3\pi \left(\rout_i\right)^3}.
\end{equation}

Now we would like to replace the star with a shell collapsing onto a
preexisting black hole.  We excise a ball of mass $m$ from the center
of the star and replace it with a mass $m$ Schwarzschild black hole.
To preserve the FRW metric in the shell's interior, we need the
initial density of the shell to be \eqref{eq:rhoi}, where $M$ is now
the total mass of the initial black hole and the shell.  In other
words, we require $m=M\left(1-\left(\rin_i/\rout_i\right)^3\right)$,
where $\rin_i$ is the initial radius of the inner surface of the
shell.  This leaves one dimensionless free parameter in the solution,
$\rin_i/\rout_i$.  The FRW metric matches to the external
Schwarzschild geometry (with mass $M$) at $\chiout$, as before, and to
the internal Schwarzschild geometry (with mass $m$) at the inner
surface of the shell at $\chiin=\sin^{-1}(\rin_i/a_i)$.  The FRW
radial coordinate is limited to the range $\chiin\leq\chi\leq\chiout$.

\subsection{Collapse solution in coordinates}

\begin{figure}
\includegraphics[width=3in,clip]{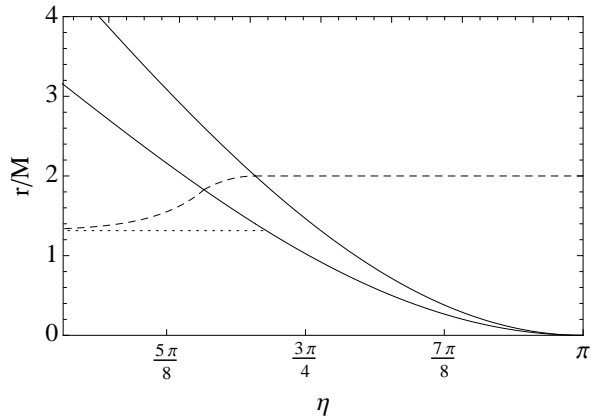}
\caption{Collapse solution as a function of the time-like coordinate
$\eta$ for $M=1$, $m=0.657$, $\rout_i=10$, and $\rin_i=7$.  The inner
and outer surfaces of the shell (solid curves) collapse to a
singularity at $\eta=\pi$.  The event horizon (dashed) approaches
$r=2m$ in the infinite past and grows to $r=2M$ as the shell
collapses.  Note that the horizon begins expanding before any material
has crossed $r=2m$ (dotted).}
\label{fig:shell}
\end{figure}

The radial coordinates of particles falling with the
shell are related to  FRW coordinates 
by:
\begin{equation}
r = \left(r_i/2\right)\left(1+\cos\eta\right).\label{eq:outerr}\\
\end{equation}
In these coordinates, every particle in the shell hits the singularity at
$\eta=\pi$.  Figure \ref{fig:shell} shows an example with $\rout_i=10$
and $\rin_i=7$.

The Schwarzschild time coordinate in the region outside the outer
boundary of the shell, $t$, is related to $\eta$ by \cite{thorne67}:
\begin{align}
t &= 2M \log
\left[
\frac{\left(\rout_i/2M-1\right)^{1/2}+\tan\left(\eta/2\right)}
     {\left(\rout_i/2M-1\right)^{1/2}-\tan\left(\eta/2\right)}
\right]\notag\\
&+2M\left(\rout_i/2M-1\right)^{1/2}
 \left[\eta+\left(\rout_i/4M\right)\left(\eta+\sin\eta\right)\right].\label{eq:outert}
\end{align}
Equations \eqref{eq:outerr} and \eqref{eq:outert}, evaluated at
$r=\rout$, describe radial free fall in the mass $M$ Schwarzschild
metric.  

The Schwarzschild time coordinate in the cavity between the initial
black hole and the shell, $t'$, is not the same as $t$:
\begin{align}
t' &= 2m \log
\left[
\frac{\left(\rin_i/2m-1\right)^{1/2}+\tan\left(\eta/2\right)}
     {\left(\rin_i/2m-1\right)^{1/2}-\tan\left(\eta/2\right)}
\right]\notag\\
&+2m\left(\rin_i/2m-1\right)^{1/2}
 \left[\eta+\left(\rin_i/4m\right)\left(\eta+\sin\eta\right)\right].\label{eq:innert}
\end{align}
The inner edge of the shell falls freely in the mass $m$
Schwarzschild metric.

\subsection{The event horizon}
\label{sec:horizon}

The outer boundary of the shell collapses across $r=2M$ at 
\begin{equation}
\eta_*=\cos^{-1}\left(\frac{4M}{\rout_i}-1\right),
\end{equation}
as follows from \eqref{eq:outerr}.  The event horizon is generated by
the null geodesic separating events which can send signals to infinity
from those that cannot. At late times, $\eta\geq\eta_*$, the horizon
is fixed at $r=2M$.  The location of the event horizon at earlier times can
be found by tracing this geodesic backwards \footnote{LZ09
mislabel this surface as the ``apparent horizon.''  The
outermost apparent horizon is at $r=2m$ for $\eta<\eta_*$ and it is at
$r=2M$ for $\eta>\eta_*$.}.

In the FRW patch in the interior of the shell, outgoing radial null
geodesics ($ds^2=d\Omega=0$) satisfy:
\begin{equation}
\frac{d\chi}{d\eta}=+1,
\end{equation}
so the solution for the horizon position in the interior of the shell is
\begin{equation}\label{eq:horizonshell}
\chi_H=\sin^{-1}\left(\rout_i/a_i\right)-\left(\eta_*-\eta\right).
\end{equation}

The horizon meets the inner boundary of the shell at
\begin{align}
\eta_{**} &= \eta_*-\Delta\eta,\\ \Delta\eta &\equiv
\sin^{-1}\left(\rout_i/a_i\right)-\sin^{-1}\left(\rin_i/a_i\right).\label{eq:deltaeta}
\end{align}
It can be extended into the infinite past according to
\begin{equation}
\frac{dr}{dt'}=1-\frac{2m}{r},
\end{equation}
which is the null geodesic equation in the mass $m$ Schwarzschild
geometry governing the cavity between the initial black hole and shell.  The
horizon position, $r_H(\eta)$, is given implicitly by:
\begin{align}\label{eq:horizoncavity}
t'(\eta)    &=r_H+2m\log(r_H-2m)+\mathcal{C},\\
\mathcal{C} &\equiv t'(\eta_{**})-(\rin(\eta_{**})+2m\log(\rin(\eta_{**})-2m)).
\end{align}
In this equation, $t'(\eta)$ is given by \eqref{eq:innert}.  To
summarize: at late times, the horizon is fixed at $r=2M$ outside the
outer surface of the shell.  At intermediate times, it expands in the
interior of the shell according to \eqref{eq:horizonshell} and, at early times, it
expands in the cavity between the black hole and shell according to
\eqref{eq:horizoncavity}.

Figure 1 shows the expanding horizon of a collapsing shell.  Note that
the horizon begins expanding before any material has crossed $r=2m$,
and it is at $r>2m$ in the cavity between the black hole
and the shell, even though the geometry is Schwarzschild
with mass $m$ there.  This illustrates the fact that event horizons depend
on the complete history of a spacetime and cannot be detected by local
measurements.  This is the teleological property of horizons.

Finally, consider the time, $t_c$, that it takes
photons emitted at the inner surface of the shell at time $\eta_e$ to
cross the shell:
\begin{equation}
t_c=t(\eta_e+\Delta \eta )-t(\eta_e).
\end{equation}
As the point of emission approaches the horizon, $\eta_e\rightarrow
\eta_{**}$, the crossing time $t_c\rightarrow t(\eta_*)-t(\eta_{**})$.  The first term 
corresponds to $r\rightarrow 2M$ in \eqref{eq:outert}, which diverges.
So an observer at infinity never sees any part of the shell
disappear behind the horizon.

\subsection{Apparent motion of the outer surface}

So far we have only considered radially emitted photons.  A photon
reaching the observer appears a distance $D=u_\phi/u_t$ from the
center of the shell, where $u_\mu$ is the photon's 4-momentum and $D$
is its impact parameter.  Radially emitted photons have $D=0$ and
appear at the center of the collapsing shell.  The apparent radius of
the shell is the largest impact parameter of the arriving photons.  
So we need to consider non-radial photon trajectories.

At each point on the shell, photons are emitted with a range of impact
parameters.  The photon with the largest impact parameter follows the
null geodesic of the Schwarzschild metric that is tangent to the
surface of the shell.  This photon has 
$u_\mu u^\mu=g^{\phi\phi}\left(u_\phi\right)^2+g^{tt}\left(u_{t}\right)^2=0$,
or
\begin{equation}\label{eq:impactouter}
D=\sqrt{\frac{g_{\phi\phi}}{-g_{tt}}}=\frac{\rout}{\sqrt{1-2M/\rout}}.
\end{equation}
This photon escapes to infinity if $\rout\geq 3M$ \cite{chandraBH}.  
We do not consider the case $\rout\leq3M$ here because we are going to find that the
apparent size of the outer surface freezes as $\rout\rightarrow 3M$. 

When $\rout\gg 3M$, we find $D\approx \rout$, the usual flat space result.  
As the shell approaches $3M$, its apparent size is given by
\eqref{eq:impactouter}.  Spacetime curvature increases the time
delay between the emission and arrival of photons, but for simplicity
we can ignore this in the early stages of collapse and use the $\rout$-motion given
by \eqref{eq:outerr}-\eqref{eq:outert}.

Geometric time delay begins to dominate the appearance of the
collapse near
\begin{equation}
D_c\equiv3\sqrt{3}M\approx 5.196M.
\end{equation}
Photons with $D\approx D_c$ spend a long period of coordinate
time on tight, outgoing spirals before escaping to infinity.  As the
outer surface of the shell passes $r=3M$, it leaves behind a
`photon cloud' whose apparent size shrinks to $D_c$
asymptotically.  
Near $D_c$, the apparent collapse can be approximated by an
exponential with e-folding time $2/\left(3\sqrt{3}M\right)$
\cite{ames68}:
\begin{equation}\label{eq:asymptote}
D=D_c+\left(D_1-D_c\right)\exp\left(-2\frac{t-t_1}{3\sqrt{3}M}\right),
\quad t>t_1,
\end{equation}
where $D_1$ is a constant and $t_1$ is the coordinate time at which
the apparent size of the shell is $D_1$.  This formula is only
strictly valid near the critical impact parameter, $D_1\rightarrow
D_c$.  To an observer at infinity, the outer surface of the shell
appears to freeze as it approaches $D_c$.

In the early stages of collapse we ignore geometrical time delay and
use \eqref{eq:impactouter} with $\rout$ given by
\eqref{eq:outerr}-\eqref{eq:outert}.  But in the late stages of
collapse, we will use \eqref{eq:asymptote} with
\begin{equation}
D_1=D_c+648 \sqrt{3}
\frac{\left(\sqrt{3}-1\right)^2}
     {\left(\sqrt{3}+1\right)^2}
\exp\left(-3 \pi\right) M
\approx 5.203M,
\end{equation}
which is the largest impact parameter for which a photon can complete
a single loop around the black hole \cite{chandraBH}.  

Figure 2 shows how the apparent size of the outer surface
asymptotically shrinks to $\dcrit$ as a shell collapses.

\begin{figure}
\includegraphics[width=3in,clip]{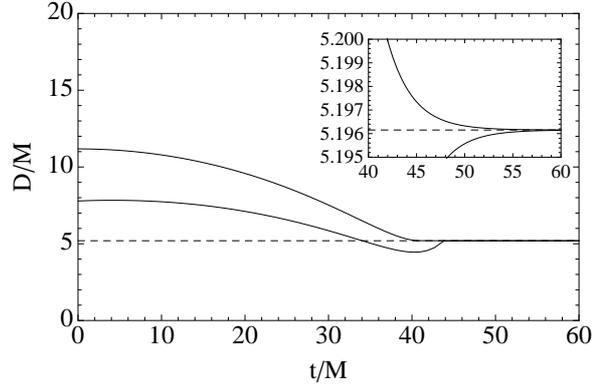}
\caption{Apparent sizes of the inner and outer surfaces of a
  collapsing shell (solid curves) as a function of the
  observer at infinity's proper time, $t$.  The collapse parameters
  are as in
  Figure 1.  At intermediate times, the apparent size of the inner
  surface varies non-monotonically.  At late times, both surfaces
  approach $\dcrit$ asymptotically (dashed).  }
\label{fig:horizon}
\end{figure}

\subsection{Apparent motion of the inner surface}

Now consider photons which leave the inner surface of the shell and
propagate outwards.  As they pass through the shell they follow
geodesics of the FRW metric.  When they reach the outer surface of the
shell they enter the external Schwarzschild geometry and continue
out to infinity.  We are going to find it convenient to replace the FRW time
coordinate $\eta$ with a comoving time coordinate $\tau$ defined by
$d\tau^2=a^2d\eta^2$.  Now $g_{\tau\tau}=-1$, and photon geodesics
satisfy \cite{weinberg}
\begin{equation}\label{eq:utau}
u_\tau=-u^\tau=a.
\end{equation}

$D$ is not conserved as photons travel across the shell because the
FRW metric \eqref{eq:frw} is time-dependent.  Let $p_1$ be the
spacetime event at which a photon is emitted at the inner surface of
the shell and let $p_2$ be the spacetime event at which it arrives at
the outer surface.  We need
\begin{equation}\label{eq:impactinner}
D=\frac{u_\phi|_{p_2}}{u_t|_{p_2}},
\end{equation}
the impact parameter of the photon as it leaves the shell.  This 
will be the impact parameter of the photon at infinity.

Matching coordinates across the outer boundary of the shell gives
\begin{equation}
u_\phi|_{p_2}=\left(\frac{\rout}{a \sin{\chi}}\right)_{p_2}u_{\tilde{\phi}} |_{p_1},
\end{equation}
where we have used the fact that $u_{\tilde{\phi}}$ is conserved along
FRW geodesics to set $u_{\tilde{\phi}}|_{p_2}=u_{\tilde{\phi}}|_{p_1}$.
We maximize $u_{\tilde{\phi}}|_{p_1}$ by considering the
photon emitted tangentially at $p_1$.  This gives
$u_{\tilde{\phi}}|_{p_1}=\sqrt{g_{\tilde{\phi}\tilde{\phi}}}u^\tau$ or 
\begin{equation}\label{eq:uphiin}
u_\phi|_{p_2}=\left(\frac{\rout}{a \sin{\chi}}\right)_{p_2}\left(\sin\chi a^2\right)_{p_1}.
\end{equation}
The energy at infinity is
\begin{equation}\label{eq:utin}
u_t|_{p_2}=\sqrt{1-2M/\rout}u_\tau|_{p_2}=\sqrt{1-2M/\rout}a.
\end{equation}
In the first step we matched coordinates across the outer boundary of
the shell and in the second step we used \eqref{eq:utau}.

Combining \eqref{eq:impactinner}, \eqref{eq:uphiin}, and
\eqref{eq:utin} gives the apparent size of the inner edge of the shell:
\begin{equation}\label{eq:impactinner2}
D=\frac{\rin(\eta-\Delta\eta)}{\sqrt{1-2M/\rout(\eta)}}.
\end{equation}
The shell crossing time, $\Delta\eta$, should be evaluated for 
geodesics emitted tangentially to the inner surface of the
shell.  However, for simplicity we will continue to use the radial time delay
\eqref{eq:deltaeta}.

In the limit of vanishing shell thickness, $\Delta\eta\rightarrow 0$,
equation \eqref{eq:impactinner2} reduces to \eqref{eq:impactouter} and
the apparent sizes of the inner and outer surfaces of the shell are
the same.  

Figure 2 shows the non-monotonicity of the apparent
size of the inner surface of the shell as it approaches $\dcrit$.

\section{Conclusions}

None of these results will be observable in the foreseeable future, as
the intensity of a star collapsing to a black hole is attenuated
exponentially on a timescale of order the black hole light crossing
time \cite{zel64}.  However, the analysis has clarified several
features of gravitational collapse and horizons.

The observer at infinity does not see any
part of the shell cross the event horizon.  It freezes entirely
outside.  We have checked that the light-crossing time across the
shell diverges as the shell approaches the horizon.  This was 
shown for the coordinates constructed in this paper and we have 
checked it is true also for the coordinates used by LZ09.

We investigated the apparent sizes of the inner and
outer surfaces of the shell during collapse.  During intermediate
stages, the apparent size of the inner surface varies
non-monotonically, an effect which is absent from the collapse of an
infinitesimally thin shell, or from the apparent collapse of the surface
of a star.  The apparent sizes of the inner and
outer surfaces of our shell both asymptote to $\dcrit$ at late times.

We fixed $m/M = 1-\left(\rin/\rout\right)^3$ for convenience, so that the
metric inside the shell would be FRW.  However, the qualitative
results are valid more generally, for generic values of $m/M$.  Our
analysis applies equally well to intermediate layers in the shell or
in the collapse of a homogeneous star.

The collapse of multiple shells onto a preexisting black hole is
qualitatively similar.  Birkhoff's theorem implies the Schwarzschild
metric is valid outside the shells.  For certain choices of the
shells' densities, FRW patches describe the spacetime inside the
shells.  The horizon now depends on the number of shells and their
relative sizes, but it can be located by tracing backwards from the
outermost shell as in \S\ref{sec:horizon}.  All of the infalling
material asymptotes to $3\sqrt{3}M$, where $M$ is now the total mass
of the initial black hole and all of the shells.  The approach to
$\dcrit$ is not monotonic for the inner layers.  An observer at
infinity sees everything freeze outside the horizon.

Our discussion has been entirely classical.  A quantum mechanical
treatment of spherical collapse is possible when the black hole mass
is much larger than the Planck mass and one again finds that
collapsing material does not appear to cross the horizon in finite
time \cite{vsk}.  This rule might be violated by Planck mass black
holes, for which quantum fluctuations of the metric are important
\cite{frolov}.  However a consideration of the black hole information
problem suggests a deeper, black hole ``complementarity'' principle
may be at work, and complete information about collapse always remains
available outside horizons \cite{thooft,susskind}.

\bibliographystyle{elsarticle-num}
\bibliography{ms.bib}

\end{document}